\documentclass[conference]{IEEEtran}
\IEEEoverridecommandlockouts

\usepackage{tikz}
\usepackage{circuitikz}
\usepackage{pgfplots} 
\usepackage{graphicx}
\usepackage{amsmath, amssymb}
\usepackage{cite}
\usepackage{float}
\usepackage{subcaption}
\usepackage{booktabs} 

\usepackage{caption}
\usepackage{siunitx}
\usepackage[hidelinks]{hyperref}
\usepackage{url}

\DeclareMathOperator*{\argmin}{\mathrm{argmin}}

\usetikzlibrary{arrows.meta, shapes.geometric, positioning}
\usetikzlibrary{calc}
\usepackage{xcolor}
\usetikzlibrary{fit, backgrounds} 

\definecolor{blockblue}{RGB}{170, 210, 245}
\definecolor{blockgreen}{RGB}{144, 238, 144}  
\definecolor{blockorange}{RGB}{255, 204, 153} 
\definecolor{blockpurple}{RGB}{216, 191, 216} 
\definecolor{blockred}{RGB}{214, 39, 40}    
\definecolor{blockgray}{RGB}{224, 224, 224}   
\definecolor{blockgrayer}{RGB}{175, 175, 175}   
\definecolor{backwarmgray}{RGB}{235,230,225}  

\colorlet{blockblue}{blockblue!60!blue}
\colorlet{blockred}{blockred!60!red}
\colorlet{blockorange}{blockorange!60!orange}

\tikzset{
    ampe/.style = {
        draw=black,
        line width=0.7pt,
        regular polygon,
        regular polygon sides=3,
        rounded corners=1pt,
        minimum size=1cm,
        shape border rotate=270,
        inner sep=-2pt
    },
    block/.style = {
        draw=black, 
        line width=0.7pt,
        rounded corners=1pt,
        minimum width=0.7cm,
        minimum height=1.2cm,
        align=center
    },
    blockclear/.style = {
        rounded corners=1pt,
        minimum width=0.7cm,
        minimum height=0.7cm,
        align=center
    },
    mixercross/.style={
    draw,
    circle,
    minimum size=0.7cm,
    line width=0.7pt,
    inner sep=0pt,
    path picture={
      \draw[line width=1pt]
        (path picture bounding box.south west) -- (path picture bounding box.north east)
        (path picture bounding box.south east) -- (path picture bounding box.north west);
        }
    },
    add/.style={
    draw,
    circle,
    minimum size=0.5cm,
    line width=0.7pt,
    inner sep=0pt,
    path picture={
      \draw[line width=1pt]
        (path picture bounding box.south) -- (path picture bounding box.north)
        (path picture bounding box.east) -- (path picture bounding box.west);
        }
    },
    blockblue/.style={block, fill=blockblue},
    blockgreen/.style={block, fill=blockgreen},
    blockorange/.style={block, fill=blockorange},
    blockpurple/.style={block, fill=blockpurple},
    blockred/.style={block, fill=blockred},
    blockgray/.style={block, fill=blockgray},
    blockgrayer/.style={block, fill=blockgrayer},
    arrow/.style = {->, line width=0.7pt},
    lightarrow/.style = {->, line width=0.7pt,draw=blockgrayer},
    modelline/.style = {-, line width=0.7pt, dotted}
}

\newcommand{\blocksep}{1cm}
\pgfplotsset{compat=1.18}

\begin{document}
\title{A Fully Open-source Implementation of an Analog 8-PAM Demapper for High-speed Communications
\vspace{-1ex}
\thanks{
This work has been partially funded by the Eindhoven Hendrik Casimir Institute (EHCI) Collaborative Projects Program and by FMD-QNC No. 16ME0831 financed by Federal Ministry of Research, Technology and Space.\\ Corresponding author: a.alvarado@tue.nl}
}
\author{\IEEEauthorblockN{Mohamed Aiham Hemza\IEEEauthorrefmark{1}, Alex Alvarado\IEEEauthorrefmark{1}, Krzysztof Herman\IEEEauthorrefmark{2}, Piyush Kaul\IEEEauthorrefmark{1}}%
\IEEEauthorblockA{\IEEEauthorrefmark{1}Eindhoven University of Technology, Eindhoven, the Netherlands}%
\IEEEauthorblockA{\IEEEauthorrefmark{2}IHP - Leibniz-Institut für innovative Mikroelektronik, Frankfurt (Oder), Germany}%
}

\maketitle

\begin{abstract}
Spectrally-efficient communication systems rely on the use of multi-level modulation formats. At the receiver side, a demodulator is often used to extract soft information about the transmitted bits. Such a demodulator is typically implemented in the digital domain. However, analog implementations of such demodulators are also possible. 
In this paper, we design and simulate an analog 8-ary pulse-amplitude modulation (8-PAM) demapper in IHP SG13G2 SiGe BiCMOS technology. We generalize and improve a design available in the literature for 4-PAM. A fully MOSFET-based 8-PAM design is proposed. Our simulations and design are completely based on open-source IC design tools. Our results show an energy efficiency of $0.33~\si{\pico\joule/\bit}$ for a data rate of $1~\si{\giga\bit/\second}$.
\end{abstract}


\section{Introduction}

Modern communication systems rely on spectrally-efficient modulation formats, such as quadrature amplitude modulation (QAM). These  formats are combined with forward error correction (FEC) to achieve high data rates. In such systems, the receiver typically provides soft information to the FEC decoder in the form of log-likelihood ratios (LLRs). This architecture is known as bit-interleaved coded modulation \cite{zehavi,caire,bicm}.

In conventional receivers, an analog-to-digital converter (ADC) is used to enable LLR computations in the digital domain. These LLRs are then processed by the FEC decoder, which also operates in the digital domain. As data rates continue to grow, the power consumption of digital circuits is rapidly growing, making it increasingly difficult to maintain a low energy per bit. Technology scaling has traditionally reduced power consumption, however, recent scaling trends have limited these gains. These challenges motivate research into receiver frontends that generate LLRs \textit{directly in the analog domain}. The ultimate goal is to improve energy efficiency and speed by adopting architectures that do not rely on ADCs.

Analog FEC decoders received considerable attention in the late 1990's and early 2000's \cite{lustenberger1999,moerz2000,xotta2002,ldpcdecoder2,morz2007analog}. These works can be traced back to the seminal work of Loeliger and Hagenauer \cite{Hagenauer1997_AEU,Loeliger2001_TIT} who recognized that the processing required in FEC decoders can be implemented using analog electronics. On the other hand, analog demappers have rarely been studied in the literature. This lack of results is slightly surprising, as analog demappers are arguably essential for implementing a fully analog receiver, which will be based on a concatenation of an analog demapper and an analog FEC decoder. 

The first analog 16-QAM demapper was proposed in \cite{seguin_2004}. The design in  \cite{seguin_2004} focuses on the constituent 4-PAM constellation and uses a PMOS differential-pair with BJT current-mirror loads to compute \textit{approximated} LLRs. These LLRs are piece-wise linear functions of the channel observation and are obtained when the max-log approximation is used \cite[Sec.~3.3.3]{bicm}. The circuit in \cite{seguin_2004} was simulated in a $0.25~\si{\micro\meter}$ BiCMOS process and relies on a hybrid design combining MOSFETs and BJTs. Two years later, \cite{frey_2006} presented an analog circuit that computes the \textit{exact} LLRs by exploiting the exponential current characteristics of MOSFETs in weak inversion. The results in\cite{seguin_2004,frey_2006} are limited to 4-PAM. Furthermore, no performance metrics (such as demapping speed or energy efficiency) were reported in \cite{seguin_2004,frey_2006}. 

In this paper, we revisit the approach of \cite{seguin_2004} and extend it to realize an 8-PAM analog demapper. Unlike the hybrid implementation in \cite{seguin_2004}, our proposed design is fully based on MOSFETs. This choice is shown to offer improved speed and lower power consumption for the same circuit configuration. The design is implemented using the IHP-Open-PDK \cite{herman2024versatility}\footnote{Recent open-source initiatives by SkyWater, GlobalFoundries and IHP have made chip design and fabrication more accessible, while the associated tools such as Qucs-S, ngspice and OpenVAF, used throughout this work, enable reliable and transparent circuit design within an open-source framework.}, which provides the SG13G2 SiGe BiCMOS technology. This process integrates HBTs operating in the sub-terahertz range on top of a fully functional CMOS platform.


\section{System Model and LLR Computations}\label{sec.systemmodel}

\begin{figure*}
    \centering
    \resizebox{0.8\linewidth}{!}{\begin{tikzpicture}

\node (b) {};
\coordinate (b1) at ($(b)+(0,0.4)$);
\coordinate (b2) at ($(b)+(0,0)$);
\coordinate (b3) at ($(b)+(0,-0.4)$);

\node[block,right=0.8\blocksep of b,fill=blockgray] (map) {Mapper\\$\mathcal{M}$};

\node[add,right=\blocksep of map,fill=blockgray] (add) {};

\node[above=0.7*\blocksep of add] (noise) {$n$};

\draw[arrow,color=blockblue] (b1) -- node[pos=0.3, above,inner sep=0] {$b_1$}  (map.west |- b1);
\draw[arrow,color=blockred] (b2) -- node[pos=0.3, above,inner sep=0] {$b_2$} (map.west |- b2);
\draw[arrow,color=blockorange] (b3) -- node[pos=0.3, above,inner sep=0] {$b_3$}  (map.west |- b3);

\node[blockgray, right=0.8\blocksep of add] (demap) {$\mathcal{M}^{-1}_{\text{d}}$};

\node[right=0.8\blocksep of demap] (dllr) {};
\coordinate (dllr1) at ($(dllr)+(0,0.4)$);
\coordinate (dllr2) at ($(dllr)+(0,0)$);
\coordinate (dllr3) at ($(dllr)+(0,-0.4)$);

\draw[arrow,color=blockblue] (demap.east |- dllr1) -- node[pos=0.65, above,inner sep=0] {$L_1$} (dllr1);
\draw[arrow,color=blockred] (demap.east |- dllr2) -- node[pos=0.65, above,inner sep=0] {$L_2$} (dllr2);
\draw[arrow,color=blockorange] (demap.east |- dllr3) -- node[pos=0.65, above,inner sep=0] {$L_3$} (dllr3);

\draw[arrow] (map) -- (add);
\draw[arrow] (map) -- node[midway, above]{$x$} (add);
\draw[arrow,-] (add) -- node[midway, above]{$r$} (demap);

\draw[arrow] (add) -- (demap);
\draw[arrow] (noise) -- (add);


\coordinate[right=3*\blocksep of demap,  yshift=0.5cm] (division);
\coordinate[right=0.05*\blocksep of division] (division_down);

\node[block, inner sep = 0.2cm, align = center, right=0 of division_down,fill=blockgray] (rtov)
{$r \to V_\text{in}$\\$V_\text{in} = \alpha r+\beta$};
\node[below=0.5pt of rtov, align = center] (textrtov) {$r$ to $V_\text{in}$\\ conversion};

\node[block, right=\blocksep of rtov, fill=blockgray] (a_demap) {$\mathcal{M}^{-1}_{\text{a}}$};

\node[block, inner sep = 0.2cm, align = center, right=1.5*\blocksep of a_demap,fill=blockgray] (vtollr)
{$V_{\text{out},{k}} \to {L}_k$\\${L}_k = \gamma_k V_{\text{out},k} + \zeta_k$};
\node[below=-2pt of vtollr, align = center] (textvtollr) {$V_{\text{out},k}$ to ${L}_k$\\ conversion};

\coordinate (allr1) at ($(vtollr.west)+(0,0.45)$);
\coordinate (allr2) at ($(vtollr.west)+(0,0)$);
\coordinate (allr3) at ($(vtollr.west)+(0,-0.45)$);

\draw[arrow,color=blockblue] (a_demap.east |- allr1) -- node[pos=0.6, above,inner sep=0] {$V_{\text{out},1}$} (allr1);
\draw[arrow,color=blockred] (a_demap.east |- allr2) -- node[pos=0.6, above,inner sep=0] {$V_{\text{out},2}$} (allr2);
\draw[arrow,color=blockorange] (a_demap.east |- allr3) -- node[pos=0.6, above,inner sep=0] {$V_{\text{out},3}$} (allr3);

\draw[arrow] (rtov) -- node[midway, above]{$V_\text{in}$} (a_demap);

\coordinate (analogboxnorthwest) at ($(rtov.north west)+(-8pt,8pt)$);
\coordinate (analogboxnortheast) at ($(vtollr.north east)+(8pt,8pt)$);
\begin{scope}[on background layer]
    \node[block,
    dashed,
    fill=backwarmgray,
    fill opacity=0.5,
    inner xsep=0pt,
    inner ysep=0pt,
    fit=(analogboxnorthwest) (textvtollr) (analogboxnortheast),
    label=above:{\small{Analog Simulation Model}}
    ] (analogbox) {};
\end{scope}

\begin{scope}[on background layer]
    \node[draw,
    rounded corners=1pt,
    dashed,
    inner sep=6pt,
    fit=(demap),
    fill=backwarmgray,
    fill opacity=0.5,
    ] (demapbox) {};
\end{scope}

\draw[arrow,-,dashed] (demapbox.north east) -- (analogbox.north west);

\draw[arrow,-,dashed] (demapbox.south east) -- (analogbox.south west);

\node[left=8pt of rtov] (inpt) {};
\draw[arrow] (inpt) -- (rtov);

\coordinate (lhat1) at ($(vtollr.east)+(8pt,0.45)$);
\coordinate (lhat2) at ($(vtollr.east)+(8pt,0)$);
\coordinate (lhat3) at ($(vtollr.east)+(8pt,-0.45)$);

\draw[arrow,-,color=blockblue] (vtollr.east |- lhat1) -- (lhat1);
\draw[arrow,-,color=blockred] (vtollr.east |- lhat2) -- (lhat2);
\draw[arrow,-,color=blockorange] (vtollr.east |- lhat3) -- (lhat3);

\end{tikzpicture}

}
    \caption{System model of the considered 8-PAM transmission. Bits $b_{1},b_{2},b_{3}$ are mapped to constellation symbols, transmitted over the channel, and converted into LLRs $L_{1},L_{2},L_{3}$ by the demapper. The r.h.s. of this figure shows the simulation model used for the proposed analog implementation of the demapper.}
    \label{fig:typicalsimulation}
\end{figure*}
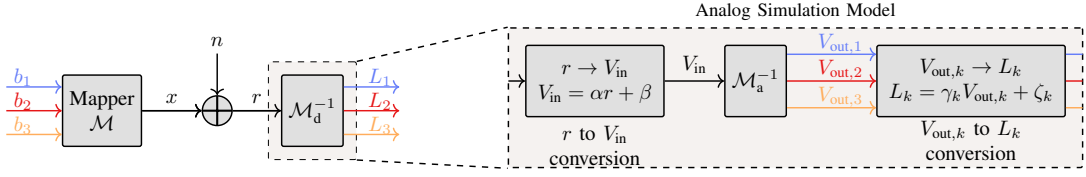

We consider communication over the additive white Gaussian noise channel. A typical baseband receiver architecture for such a system includes an ADC followed by different digital signal processing subsystems (e.g., equalization, demapping, and decoding). We consider 8-PAM with equally spaced symbols, thereby effectively considering 64-QAM. The transmitted bits $b_1,b_2,b_3$ are assumed to be independent and uniformly distributed. The transmitted symbol $x$ is modeled as a discrete random variable, uniformly distributed over the 8-PAM constellation, i.e., $X \sim \mathcal{U}(\mathcal{A})$, where 
    $\mathcal{A} = \{\pm 7d, \pm 5d, \pm 3d, \pm 1d\}$. 
The scaling factor $d$ is chosen so that the average symbol energy is
    $\mathbb{E}[X^2] = 0.5$, and thus, the QAM constellation has unit energy. The bit-to-symbol mapping ($\mathcal{M}$) is based on the binary-reflected Gray code \cite{Agrell04}.

At every discrete-time instant, the received signal is 
    $r = x + n$,
where the noise $N\sim \mathcal{N}(0, \sigma^2)$ is Gaussian. In a traditional  receiver, the received signal $r$ represents the digital signal that is fed to a digital demapper $\mathcal{M}^{-1}_d$, which produces the LLRs $L$. This configuration is shown in Fig.~\ref{fig:typicalsimulation} (left) for the case of 8-PAM, where $3$ LLRs are computed for each $r$.

An LLR gives a soft estimate of how likely a transmitted bit is ($1$ or $0$), given the received symbol \cite{bicm}. It is defined as 
\begin{equation}\label{llr}
    L_k \triangleq \log \frac{p_{R|B_k}(r|1)}{p_{R|B_k}(r|0)},
\end{equation}
where $B_k$ is the $k$-th bit in the symbol and $p_{R|B_k}(r | b)$ is the likelihood of receiving $r$ assuming $B_k=b$.  For the channel under consideration, the LLR in \eqref{llr} is 
\begin{align}\label{exact}
    L_k = \log { \sum\limits_{i \in \mathcal{I}_k^1} \mathrm{e}^{-\frac{(r - x_i)^2}{2\sigma^2} } }
    -\log{ \sum\limits_{i \in \mathcal{I}_k^0} \mathrm{e}^{-\frac{(r - x_i)^2}{2\sigma^2}}},
\end{align}
where $\mathcal{I}_k^b$ is the set of indices of constellation points where bit $B_k=b$ and the signal to noise ratio (SNR) is $\text{SNR}\triangleq 1/2\sigma^2$. 

The exponential functions in \eqref{exact} decay rapidly, and thus, the terms in the LLR expression are dominated by the smallest distances $(r - x_{i})^2$. Therefore, an approximation is obtained by retaining only the $x_{i}$ closest to $r$ from $\mathcal{I}_k^b$. This leads to the max-log approximation \cite[eq.~(3.99)]{bicm}
\begin{equation}\label{ml}
    L_k \approx \text{SNR} \left( \min\limits_{i \in \mathcal{I}_k^0} (r - x_i)^2 - \min\limits_{i \in \mathcal{I}_k^1} (r - x_i)^2 \right).
\end{equation}

\begin{figure}
    \centering
    
    \begin{subfigure}{\linewidth}
        \centering
        \definecolor{Cblue}{RGB}{31 119 180}
\definecolor{Cred}{RGB}{214 39 40}
\definecolor{Corange}{RGB}{255 127 14}
\definecolor{Cgreen}{RGB}{44 160 44}
\definecolor{Cpurple}{RGB}{148 103 198}
\definecolor{Cbrown}{RGB}{140 86 75}
\definecolor{Cpink}{RGB}{227 119 194}
\definecolor{Ccyan}{RGB}{23 190 207}
\definecolor{Cgray}{RGB}{127 127 127}
\definecolor{Cgray}{RGB}{0 0 0}
\definecolor{Cyellow}{RGB}{188 189 34}

\begin{tikzpicture}
\begin{axis}[
grid=both,
    width=1.07\columnwidth, height=4.5cm,
    axis x line=middle,
    axis y line=middle,
    axis line style={-{Latex[length=3pt,width=3pt]}, thick},
    xmin=0, xmax=0.64,
    xtick={0.04, 0.12, 0.20, 0.28},
    extra x ticks={0.36,0.44,0.52,0.60},
    xticklabel style={
        /pgf/number format/fixed,
        /pgf/number format/precision=2,
        /pgf/number format/fixed zerofill,
        font=\scriptsize
    },
    extra x tick style={
        ticklabel style={anchor=south},
        major tick length=2pt,
    },
    enlargelimits=false,
    clip=false,
    legend columns=2,
    legend style={ fill=none,
        font=\footnotesize,
        fill=white,
        at={(0.65,0.67)},
        anchor=south
    },
    yticklabel style={font=\footnotesize},
    xlabel={$V_\text{in}$},
    ylabel={$L_1$},
    xlabel style={
        at={(axis description cs:1,0.5)},
        anchor=north east,
        font=\footnotesize
    },
    ytick={-15,-10,-5,5,10,15},
    yticklabels={\phantom{.}$-15$,$-10$,$-5$,$5$,$10$,$15$},
    ylabel style={
        at={(axis description cs:0,1)},
        anchor= east,
        font=\footnotesize
    },
    axis background/.style={draw=black}
]

{\addplot [color=Cgray, 
thick,
mark=none, 
mark options={solid, fill=white},
every mark/.append style={scale=1},
mark repeat=100,
mark phase=40
]
table [x = r, y = LLR_k1_exact, col sep=space] {figures/LLR_results_with_qucs_r.txt};
}\addlegendentry{Digital Exact};

{\addplot [color=Cgreen, 
thick, 
mark=none, 
mark options={solid, fill=white},
every mark/.append style={scale=1},
mark repeat=100,
mark phase=65
]
table [x = r, y = LLR_k1_ML, col sep=space] {figures/LLR_results_with_qucs_r.txt};
}\addlegendentry{Digital Max-Log};


{\addplot [color=Cred, 
thick, 
mark=none, 
mark options={solid, fill=white},
every mark/.append style={scale=1},
mark repeat=101,
mark phase=12
]
table [x = Vin, y = Vout, col sep=space] {figures/LLR_k1_qucs.txt};
}\addlegendentry{Analog BJT};

{\addplot [color=Cblue, 
thick, 
mark=none, 
mark options={solid, fill=white},
every mark/.append style={scale=1},
mark repeat=101,
mark phase=12
]
table [x = Vin, y = Vout, col sep=space] {figures/LLR_MOS_k1.txt};
}\addlegendentry{Analog MOSFET};

\addplot[only marks, mark=*, mark options={scale=0.9, fill=black}] coordinates {(0.04,0)};

\addplot[only marks, mark=*, mark options={scale=0.9, fill=black}] coordinates {(0.12,0)};

\addplot[only marks, mark=*, mark options={scale=0.9, fill=black}] coordinates {(0.2,0)};

\addplot[only marks, mark=*, mark options={scale=0.9, fill=black}] coordinates {(0.28,0)};

\addplot[only marks, mark=*, mark options={scale=0.9, fill=black}] coordinates {(0.36,0)};

\addplot[only marks, mark=*, mark options={scale=0.9, fill=black}] coordinates {(0.44,0)};

\addplot[only marks, mark=*, mark options={scale=0.9, fill=black}] coordinates {(0.52,0)};

\addplot[only marks, mark=*, mark options={scale=0.9, fill=black}] coordinates {(0.6,0)};

\end{axis}


\end{tikzpicture}
        \label{llr1}
    \end{subfigure}
    \begin{subfigure}{\linewidth}
        \centering
        \vspace{-2ex}
        \definecolor{Cblue}{RGB}{31 119 180}
\definecolor{Cred}{RGB}{214 39 40}
\definecolor{Corange}{RGB}{255 127 14}
\definecolor{Cgreen}{RGB}{44 160 44}
\definecolor{Cpurple}{RGB}{148 103 198}
\definecolor{Cbrown}{RGB}{140 86 75}
\definecolor{Cpink}{RGB}{227 119 194}
\definecolor{Ccyan}{RGB}{23 190 207}
\definecolor{Cgray}{RGB}{127 127 127}
\definecolor{Cgray}{RGB}{0 0 0}
\definecolor{Cyellow}{RGB}{188 189 34}

\begin{tikzpicture}
\begin{axis}[
grid=both,
    width=1.07\columnwidth, height=4.5cm,
    axis x line=middle,
    axis y line=middle,
    axis line style={-{Latex[length=3pt,width=3pt]}, thick},
    xmin=0, xmax=0.64, ymin=-6.4,
    xtick={0.04, 0.12, 0.52, 0.60},
    extra x ticks = {0.20, 0.28, 0.36, 0.44},
    ytick={-4.5,-3,-1.5,1.5,3,4.5},
    xticklabel style={
        /pgf/number format/fixed,
        /pgf/number format/precision=2,
        /pgf/number format/fixed zerofill,
        font=\scriptsize
    },
    extra x tick style={
        ticklabel style={anchor=south},
        major tick length=2pt,
    },
    yticklabel style={font=\footnotesize},
    enlargelimits=false,
    clip=false,
    xlabel={$V_\text{in}$},
    ylabel={$L_2$},
    xlabel style={
        at={(axis description cs:1,0.5)},
        anchor=south east,
        font=\footnotesize
    },
    ylabel style={
        at={(axis description cs:0,1)},
        anchor=east,
        font=\footnotesize
    },
    axis background/.style={draw=black}
]

{\addplot [color=Cgray, 
thick,
mark=none, 
mark options={solid, fill=white},
every mark/.append style={scale=1},
mark repeat=100,
mark phase=40
]
table [x = r, y = LLR_k2_exact, col sep=space] {figures/LLR_results_with_qucs_r.txt};
}

{\addplot [color=Cgreen, 
thick, 
mark=none, 
mark options={solid, fill=white},
every mark/.append style={scale=1},
mark repeat=100,
mark phase=65
]
table [x = r, y = LLR_k2_ML, col sep=space] {figures/LLR_results_with_qucs_r.txt};
}


{\addplot [color=Cred, 
thick, 
mark=none, 
mark options={solid, fill=white},
every mark/.append style={scale=1},
mark repeat=101,
mark phase=12
]
table [x = Vin, y = Vout, col sep=space] {figures/LLR_k2_qucs.txt};
}

{\addplot [color=Cblue, 
thick, 
mark=none, 
mark options={solid, fill=white},
every mark/.append style={scale=1},
mark repeat=101,
mark phase=12
]
table [x = Vin, y = Vout, col sep=space] {figures/LLR_MOS_k2.txt};
}

\addplot[only marks, mark=*, mark options={scale=0.9, fill=black}] coordinates {(0.04,0)};

\addplot[only marks, mark=*, mark options={scale=0.9, fill=black}] coordinates {(0.12,0)};

\addplot[only marks, mark=*, mark options={scale=0.9, fill=black}] coordinates {(0.2,0)};

\addplot[only marks, mark=*, mark options={scale=0.9, fill=black}] coordinates {(0.28,0)};

\addplot[only marks, mark=*, mark options={scale=0.9, fill=black}] coordinates {(0.36,0)};

\addplot[only marks, mark=*, mark options={scale=0.9, fill=black}] coordinates {(0.44,0)};

\addplot[only marks, mark=*, mark options={scale=0.9, fill=black}] coordinates {(0.52,0)};

\addplot[only marks, mark=*, mark options={scale=0.9, fill=black}] coordinates {(0.6,0)};

\end{axis}
\end{tikzpicture}
        \label{fig:llr2}
    \end{subfigure}
    \begin{subfigure}{\linewidth}
        \centering
        \vspace{-2ex}
        \definecolor{Cblue}{RGB}{31 119 180}
\definecolor{Cred}{RGB}{214 39 40}
\definecolor{Corange}{RGB}{255 127 14}
\definecolor{Cgreen}{RGB}{44 160 44}
\definecolor{Cpurple}{RGB}{148 103 198}
\definecolor{Cbrown}{RGB}{140 86 75}
\definecolor{Cpink}{RGB}{227 119 194}
\definecolor{Ccyan}{RGB}{23 190 207}
\definecolor{Cgray}{RGB}{127 127 127}
\definecolor{Cgray}{RGB}{0 0 0}
\definecolor{Cyellow}{RGB}{188 189 34}

\begin{tikzpicture}
\begin{axis}[
grid=both,
    width=1.07\columnwidth, height=4.5cm,
    axis x line=middle,
    axis y line=middle,
    axis line style={-{Latex[length=3pt,width=3pt]}, thick},
    xmin=0, xmax=0.64, ymin=-2.15,
    xtick={0.04, 0.28, 0.36, 0.60},
    extra x ticks = {0.12, 0.20, 0.44, 0.52},
    ytick={-1.5, -1,-0.5,0.5, 1, 1.5},
    yticklabel style={font=\footnotesize},
    xticklabel style={
        /pgf/number format/fixed,
        /pgf/number format/precision=2,
        /pgf/number format/fixed zerofill,
        font=\scriptsize
    },
    extra x tick style={
        ticklabel style={anchor=south},
        major tick length=2pt,
    },
    enlargelimits=false,
    clip=false,
    xlabel={$V_\text{in}$},
    ylabel={$L_3$},
    xlabel style={
        at={(axis description cs:1,0.5)},
        anchor=south east,
        font=\footnotesize
    },
    ylabel style={
        at={(axis description cs:0,1)},
        anchor=east,
        font=\footnotesize
    },
    axis background/.style={draw=black}
]

{\addplot [color=Cgray, 
thick,
mark=none, 
mark options={solid, fill=white},
every mark/.append style={scale=1},
mark repeat=100,
mark phase=40
]
table [x = r, y = LLR_k3_exact, col sep=space] {figures/LLR_results_with_qucs_r.txt};
}

{\addplot [color=Cgreen, 
thick, 
mark=none, 
mark options={solid, fill=white},
every mark/.append style={scale=1},
mark repeat=100,
mark phase=65
]
table [x = r, y = LLR_k3_ML, col sep=space] {figures/LLR_results_with_qucs_r.txt};
}

{\addplot [color=Cblue, 
thick, 
mark=none, 
mark options={solid, fill=white},
every mark/.append style={scale=1},
mark repeat=100,
mark phase=65
]
table [x = Vin, y = Vout, col sep=space] {figures/LLR_MOS_k3.txt};
}


{\addplot [color=Cred, 
thick, 
mark=none, 
mark options={solid, fill=white},
every mark/.append style={scale=1},
mark repeat=101,
mark phase=12
]
table [x = Vin, y = Vout, col sep=space] {figures/LLR_k3_qucs.txt};
}

\addplot[only marks, mark=*, mark options={scale=0.9, fill=black}] coordinates {(0.04,0)};

\addplot[only marks, mark=*, mark options={scale=0.9, fill=black}] coordinates {(0.12,0)};

\addplot[only marks, mark=*, mark options={scale=0.9, fill=black}] coordinates {(0.2,0)};

\addplot[only marks, mark=*, mark options={scale=0.9, fill=black}] coordinates {(0.28,0)};

\addplot[only marks, mark=*, mark options={scale=0.9, fill=black}] coordinates {(0.36,0)};

\addplot[only marks, mark=*, mark options={scale=0.9, fill=black}] coordinates {(0.44,0)};

\addplot[only marks, mark=*, mark options={scale=0.9, fill=black}] coordinates {(0.52,0)};

\addplot[only marks, mark=*, mark options={scale=0.9, fill=black}] coordinates {(0.6,0)};


\end{axis}
\end{tikzpicture}
        \label{fig:llr3}
    \end{subfigure}
    \vspace{-5ex}
    \caption{LLRs for bit positions  $k=1$ (top), $k=2$ (middle), and  $k=3$ (bottom) at $\text{SNR}=10~\si{\decibel}$. The received symbol $r$ is scaled by $(\alpha,\beta)$ to obtain $V_{\text{in}}$. The scaled constellation points are also shown. The analog demapper output voltages $V_{\text{out}}$ are scaled by $(\gamma_k,\zeta_k)$ to obtain the LLRs.}
    \vspace{-4ex}
    \label{fig:llr}
\end{figure}
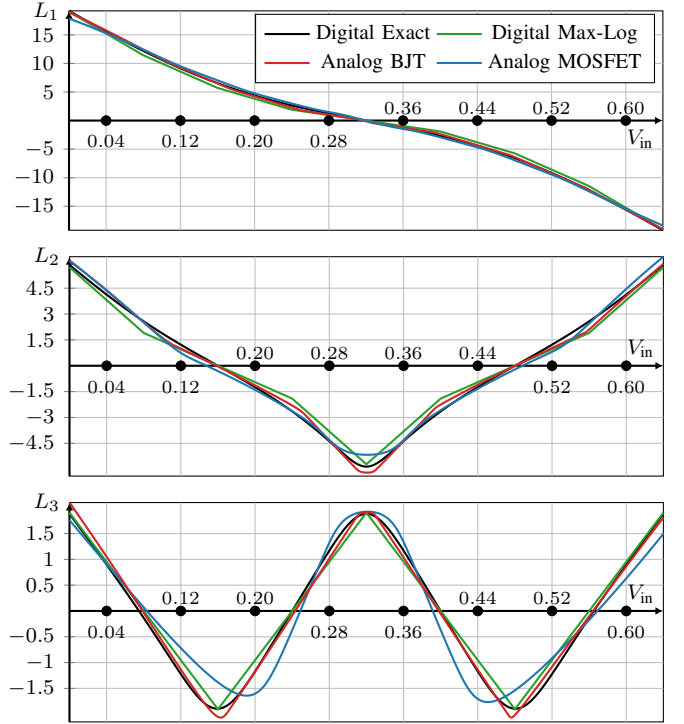

Fig.~\ref{fig:llr} shows the exact and max-log LLRs with black and green lines, resp. The x-axis shown is scaled and shifted to map the constellation points $\mathcal{A}$ to the analog demapper's input range. Details on this scaling and shifting are given in Sec.~\ref{sec.design}. The results in this figure show how the max-log approximation effectively converts the nonlinear relationship between $r$ and $L_k$ (black lines) into a piece-wise linear function (green lines). This figure also shows results obtained using analog circuits (red and blue lines), whose design is discussed next.

\section{Proposed Design}\label{sec.design}

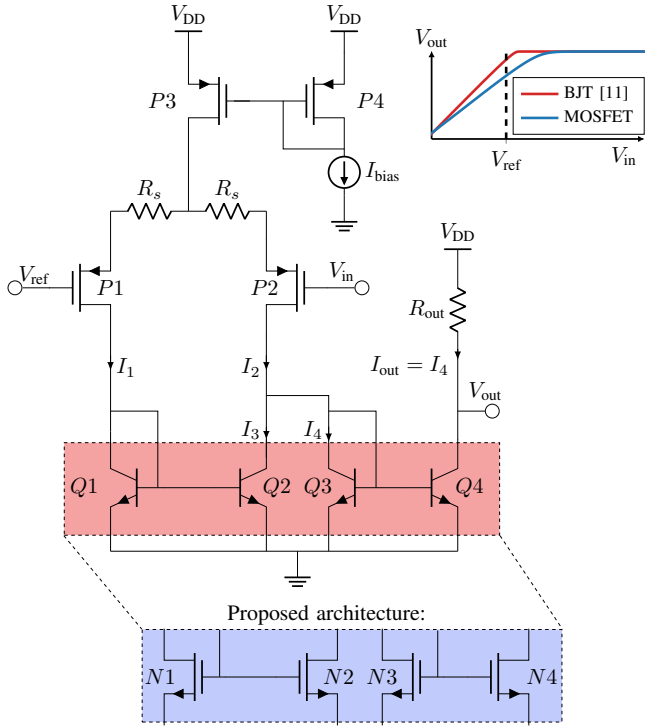
\begin{figure}[t]
    \centering
    \resizebox{0.98\linewidth}{!}{\begin{tikzpicture}
	\node[shape=rectangle, fill=blockblue, fill opacity=0.4, draw, dash pattern={on 1.6pt off 1.6pt}, minimum width=6.736cm, minimum height=1.486cm] at (7.875, 2.75){};
	\node[shape=rectangle, fill=blockred, fill opacity=0.4, draw, dash pattern={on 1.6pt off 1.6pt}, minimum width=6.986cm, minimum height=1.486cm] at (6.75, 5.75){};
	\node[npn, xscale=-1](N1) at (4, 5.77){} node[anchor=east] at (N1.text){$Q1$};
	\node[npn](N2) at (6.5, 5.77){} node[anchor=west] at ([xshift=-0.25cm]N2.text){$Q2$};
	\draw (4, 6.67) -| (4, 8.21);
	\draw (6.5, 6.25) -| (6.5, 8.21);
	\node[pmos, arrowmos, nocircle](N3) at (4, 8.98){} node[anchor=west] at ([xshift=-0.42cm]N3.text){$P1$};
	\node[pmos, arrowmos, nocircle, xscale=-1](N4) at (6.5, 8.98){} node[anchor=east] at ([xshift=0.39cm]N4.text){$P2$};
	\draw (5.25, 10.21) to[american resistor, /tikz/circuitikz/bipoles/length=0.840cm, l_={$R_s$}] (4, 10.21);
	\draw (6.5, 10.21) to[american resistor, /tikz/circuitikz/bipoles/length=0.840cm, l_={$R_s$}] (5.25, 10.21);
	\node[npn, xscale=-1](N5) at (7.5, 5.77){} node[anchor=east] at ([xshift=0.25cm]N5.text){$Q3$};
	\node[npn](N6) at (9.571, 5.77){} node[anchor=west] at ([xshift=-0.23cm]N6.text){$Q4$};
	\draw (6.5, 7.25) -| (7.5, 6.54);
	\draw (7.5, 7) -| (8.26, 5.77);
	\draw (9.571, 6.54) -| (9.571, 8);
	\draw (5.25, 10.21) -| (5.25, 11.21);
	\node[ocirc, xscale=1.99, yscale=1.99](N7) at (8.034, 8.98){} node[anchor=south east] at ([xshift=0.12cm, yshift=-0.12cm]N7.north west){$V_\text{in}$};
	\draw (7.48, 8.98) -| (7.909, 8.985);
	\node[ocirc, xscale=-1.99, yscale=1.99](N8) at (2.466, 8.98){} node[anchor=south west] at ([xshift=-0.18cm, yshift=-0.18cm]N8.north west){$V_\text{ref}$};
	\draw (2.591, 8.98) -| (3.02, 8.985);
	\draw (4, 5) -| (4, 4.75);
	\draw (6.5, 5) -| (6.5, 4.75);
	\draw (7.5, 5) -| (7.5, 4.75);
	\node[rground, yscale=-1](N9) at (5.25, 12.75){} node[anchor=south] at (N9.south){$V_\text{DD}$};
	\node[pmos, arrowmos, nocircle, xscale=-1](N10) at (5.25, 11.98){} node[anchor=east] at (N10.text){$P3$};
	\node[pmos, arrowmos, nocircle](N11) at (7.75, 11.98){} node[anchor=west] at (N11.text){$P4$};
	\draw (7.75, 11.21) to[american current source, mirror, /tikz/circuitikz/bipoles/length=0.840cm, l={$I_{\text{bias}}$}] (7.75, 10.46);
	\draw (5.98, 11.98) -- (6.77, 11.98);
	\node[rground, xscale=-1, yscale=-1](N12) at (7.75, 12.75){} node[anchor=south] at (N12.south){$V_\text{DD}$};
	\node[ground, xscale=-1] at (7.75, 10.46){};
	\node[ground] at (7, 4.75){};
	\draw (4, 10.21) -- (4, 9.75);
	\draw (6.5, 10.21) -| (6.5, 9.75);
	\draw (8.34, 5.77) -- (8.731, 5.77);
	\draw (9.571, 5) -| (9.571, 4.75);
	\draw (4, 4.75) -- (9.571, 4.75);
	\draw (6.77, 11.98) |- (7.75, 11.21);
	\node[shape=rectangle, minimum width=0.465cm, minimum height=-0.035cm](N13) at (8.797, 7.732){} node[anchor=center] at (N13.text){$I_\text{out}=I_4$} node[anchor=north west, align=left, text width=0.077cm, inner sep=6pt] at (8.547, 7.732){};
	\draw[-latex] (4, 7.897) -| (4, 7.647);
	\draw[-latex] (6.5, 6.75) -| (6.5, 6.54);
	\draw[-latex] (6.5, 7.897) -| (6.5, 7.647);
	\draw[-latex] (7.5, 6.75) -| (7.5, 6.5);
	\draw[-latex] (9.571, 8) -| (9.571, 7.817);
	\node[shape=rectangle, minimum width=0.465cm, minimum height=0.362cm](N14) at (4.25, 7.699){} node[anchor=center] at (N14.text){$I_1$} node[anchor=north west, align=left, text width=0.077cm, inner sep=6pt] at (4, 7.897){};
	\node[shape=rectangle, minimum width=0.465cm, minimum height=0.362cm](N15) at (6.25, 7.699){} node[anchor=center] at (N15.text){$I_2$} node[anchor=north west, align=left, text width=0.077cm, inner sep=6pt] at (6, 7.897){};
	\node[shape=rectangle, minimum width=0.465cm, minimum height=0.362cm](N16) at (6.25, 6.699){} node[anchor=center] at (N16.text){$I_3$} node[anchor=north west, align=left, text width=0.077cm, inner sep=6pt] at (6, 6.897){};
	\node[shape=rectangle, minimum width=0.465cm, minimum height=0.362cm](N17) at (7.25, 6.699){} node[anchor=center] at (N17.text){$I_4$} node[anchor=north west, align=left, text width=0.077cm, inner sep=6pt] at (7, 6.897){};
	\draw (9.571, 8) to[american resistor, /tikz/circuitikz/bipoles/length=0.840cm, l={$R_\text{out}$}] (9.571, 9.25);
	\node[rground, xscale=-1, yscale=-1](N18) at (9.571, 9.25){} node[anchor=south] at (N18.south){$V_\text{DD}$};
	\node[ocirc, xscale=1.99, yscale=1.99](N19) at (10.125, 7){} node[anchor=south] at ([yshift=-0.06cm]N19.north west){$V_\text{out}$};
	\draw (9.571, 7) -| (10, 7.005);
	\draw (4.84, 5.77) -- (5.66, 5.77);
	\draw (9.34, 2.73) -- (9.731, 2.73);
	\draw (5.84, 2.73) -- (6.66, 2.73);
	\node[nmos, arrowmos](N20) at (7.64, 2.73){} node[anchor=west] at ([xshift=-0.42cm]N20.text){$N2$};
	\node[nmos, arrowmos](N21) at (10.711, 2.73){} node[anchor=west] at ([xshift=-0.23cm]N21.text){$N4$};
	\node[nmos, arrowmos, xscale=-1](N22) at (8.36, 2.73){} node[anchor=east] at ([xshift=0.44cm]N22.text){$N3$};
	\node[nmos, arrowmos, xscale=-1](N23) at (4.86, 2.73){} node[anchor=east] at ([xshift=0.38cm]N23.text){$N1$};
	\draw (5.75, 3.485) -- (5.75, 2.735);
	\draw (9.25, 3.477) -- (9.25, 2.727);
	\draw (4, 6.67) -| (4, 6.54);
	\draw (4.758, 5.773) |- (4, 7);
	\draw[dash pattern={on 1.6pt off 1.6pt}] (3.25, 5) -- (4.5, 3.5);
	\draw[dash pattern={on 1.6pt off 1.6pt}] (10.25, 5) -- (11.25, 3.5);
	\node[shape=rectangle, minimum width=4.215cm, minimum height=0.465cm] at (7.786, 3.847){} node[anchor=north west, align=left, text width=3.827cm, inner sep=6pt] at (5.661, 4.097){Proposed architecture:};

    \node[anchor=south west, xshift=-4cm, yshift=-3cm] (topplot) at (current bounding box.north east) {\definecolor{Cblue}{RGB}{31 119 180}
\definecolor{Cred}{RGB}{214 39 40}
\definecolor{Corange}{RGB}{255 127 14}
\definecolor{Cgreen}{RGB}{44 160 44}
\definecolor{Cpurple}{RGB}{148 103 198}
\definecolor{Cbrown}{RGB}{140 86 75}
\definecolor{Cpink}{RGB}{227 119 194}
\definecolor{Ccyan}{RGB}{23 190 207}
\definecolor{Cgray}{RGB}{127 127 127}
\definecolor{Cgray}{RGB}{0 0 0}
\definecolor{Cyellow}{RGB}{188 189 34}

\begin{tikzpicture}
\begin{axis}[
    width=5cm, height=3cm,
    axis x line=middle,
    axis y line=middle,
    axis line style={-{Latex[length=3pt,width=3pt]}, thick},
    xmin=0.25, xmax=0.45,
    xtick={0.32},
    xticklabels={$V_\text{ref}$},
    ytick=\empty,
    legend style={ fill=none,
        font=\footnotesize,
        at={(1,0.7)},
        legend cell align=left
    },
    xlabel={$V_\text{in}$},
    ylabel={$V_\text{out}$},
    xlabel style={at={(axis description cs:1,0)}, anchor=north east},
    ylabel style={at={(axis description cs:0,0.95)}, anchor=south},
]

{\addplot [color=Cred, 
very thick, 
mark=none
]
table [x = vin, y = bjt, col sep=space] {figures/BJTMOSsweep.txt};
}\addlegendentry{BJT \cite{seguin_2004}};

{\addplot [color=Cblue, 
very thick, 
mark=none
]
table [x = vin, y = mos, col sep=space] {figures/BJTMOSsweep.txt};
}\addlegendentry{MOSFET};

\addplot [dashed, very thick] coordinates {(0.32,1.59) (0.32,1.6)};

\end{axis}
\end{tikzpicture}};



    
\end{tikzpicture}}
    \caption{Single cell implementing one segment of the piece-wise linear LLR function. The inset shows the transfer characteristics for the original design \cite{seguin_2004} (BJTs) and our proposed architecture (MOSFETs).}
    \vspace{-2ex}
    \label{fig:circuit}
\end{figure}

Here we consider using an analog demapper $\mathcal{M}_{\text{a}}^{-1}$ to replace the digital demapper $\mathcal{M}_{\text{d}}^{-1}$ in Fig.~\ref{fig:typicalsimulation}. This analog demapper is fully described by its input-output relationship, where $V_{\text{in}}$ is the input voltage, and $V_{\text{out},k}$ with $k=1,2,3$ are the output voltages. 
To implement such an analog demapper, \cite{seguin_2004} proposed using a cell-based approach, where each cell generates a piece-wise linear function. These cells are then combined to obtain the piece-wise linear relationship created by the max-log approximation in \eqref{ml}. The cell proposed in \cite{seguin_2004} is shown in Fig.~\ref{fig:circuit}. The resulting single piece-wise function is shown with a red line in the inset of Fig.~\ref{fig:circuit}.

The cell in Fig.~\ref{fig:circuit} operates as a current-steering stage that converts $V_\text{in}$ into a piece-wise linear output current. Transistors $P3$-$P4$ generate a bias current $I_\text{bias}$ which is mirrored into the PMOS differential pair $P1$-$P2$. Depending on the values of $V_{\text{in}}$ and $V_{\text{ref}}$, more current is directed toward $I_1$ or $I_2$, modulating the collector currents of $Q1$-$Q2$. $Q1$-$Q2$ mirror the current $I_1$ to $I_3$ depending on the relation between $I_1$ and $I_2$. By Kirchhoff's current law, $I_4=I_2-I_3$ which is then mirrored by $Q3$-$Q4$ translating it to $I_\text{out}$. BJTs were proposed in \cite{seguin_2004} to be used in the mirrors for improved linearity, effectively trying to mimic the max-log approximation in the best possible way. The voltage transfer characteristic can be analyzed based on three cases for $V_{\text{in}}$ and $V_{\text{ref}}$:

\noindent \textbf{Case 1} ($V_{\text{in}} < V_{\text{ref}}$): $P2$ sinks more current than $P1$, so $I_2 > I_1$. $I_1=I_3$ and $I_\text{out}=I_2-I_1$, giving $V_\text{out}=V_\text{DD}-I_\text{out}R_\text{out}$.

\noindent \textbf{Case 2} ($V_{\text{in}} = V_{\text{ref}}$): $P1$-$P2$ conduct equal current $I_2 = I_1$, therefore, $I_\text{out}=0$ and $V_\text{out}=V_\text{DD}$.

\noindent \textbf{Case 3} ($V_{\text{in}} > V_{\text{ref}}$): $P2$ sinks less current than $P1$. As a result, $Q2$ is driven into saturation (base-emitter and base-collector are forward biased). No current is mirrored through $Q1$-$Q2$, hence $I_3=I_2$. Therefore $I_\text{out}=0$ and $V_\text{out}=V_\text{DD}$.

The voltage transfer function in the inset of Fig.~\ref{fig:circuit} shows a positive slope, followed by constant voltage $V_\text{out}=V_\text{DD}$. Swapping $V_{\text{in}}$ and $V_{\text{ref}}$ generates a constant voltage $V_\text{out}=V_\text{DD}$ followed by a descending slope. The 8-PAM LLRs are then realized by adding and subtracting the output currents of multiple cells with different reference voltages. To allow this, a differential output structure is employed, where each cell contributes either to the positive or the negative output branch, resulting in
\begin{equation}
    V_{\text{out}} = V_\text{DD} - \biggl(\,\sum_{i\in \text{pos}}{I_{\text{out},i}}-\sum_{i\in \text{neg}}{I_{\text{out},i}}\biggr) \cdot R_\text{out},
\end{equation}
where $i\in \text{pos}$ ($i\in \text{neg}$) corresponds to cells whose output currents contribute positively (negatively) to the output.

While the cell proposed in \cite{seguin_2004} provides good linearity, its operation speed is limited by charge storage effects when the BJTs enter saturation. To overcome this limitation, we propose to replace $Q1$-$Q4$ with NMOS transistors $N1$-$N4$. This slightly reduced the linearity of the piecewise characteristic, as shown in the inset of Fig.~\ref{fig:circuit}. However, as we will see in Sec.~\ref{sec.results}, using NMOS enables higher operating speed and lower energy consumption. In the following sections, the circuit from \cite{seguin_2004} and the newly proposed design are referred to as the BJT and MOSFET designs, respectively.

To ensure that $P1$-$P4$ remain in saturation\footnote{Note that MOSFET saturation occurs when the channel is pinched off near the drain, and is different than BJT saturation.}, the input voltage was limited to $V_\text{in}\leq0.64~\si{\volt}$. The supply voltage was fixed to the maximum allowed by the technology, i.e., $V_\text{DD}=1.6~\si{\volt}$. The output resistance was set to $R_\text{out}=3~\si{\kilo\ohm}$. This value can be optimized depending on the subsequent stage. The source resistance $R_s=500~\si{\ohm}$ helps maintain output linearity over a wide range of input voltage swings. The transistors were left at minimum size ($0.15~\si{\micro\meter}$ width and $0.14~\si{\micro\meter}$ length for MOSFETs and 1 emitter with length $0.9~\si{\micro\meter}$ for BJTs).

\section{Results and Discussion}\label{sec.results}

To simulate the analog demapper, the input is a voltage $V_\text{in}$ rather than $r$, and the outputs $V_{\text{out},k}$ are voltages rather than a digital LLRs. To enable a fair comparison, conversion blocks are introduced, as shown in Fig.~\ref{fig:typicalsimulation}. The input voltage is mapped via
$
    V_\text{in} = \alpha r + \beta,
$
where $(\alpha,\beta)$ define the mapping between the constellation points $\mathcal{A}$ and the circuit’s input range. They were chosen such that $r=-7d$ and $r=+7d$ map to $V_\text{in}=0.04~\si{\volt}$ and $V_\text{in}=0.60~\si{\volt}$ respectively, resulting in $\alpha=0.2592$ and $\beta=0.32$. The analog outputs are then linearly transformed per bit as
$
    L_k = \gamma_k V_{\text{out},k} + \zeta_k$,
where $(\gamma_k,\zeta_k)$ are obtained by minimizing the squared amplitude error with respect to the reference LLRs, i.e.,
\begin{equation}\label{opt}
    (\gamma_k,\zeta_k)
=\argmin_{\gamma,\zeta}\;\int_{-\infty}^{\infty}(\gamma\,V_{\text{out},k}(r)+\zeta-L_k(r))^2\,\mathrm{d}r,
\end{equation}
where $L_k$ is given by \eqref{exact} and the dependency of $L_k$ and $V_{\text{out},k}$ on the channel observation $r$ was made explicit. the optimization \eqref{opt} is performed for each SNR.

The performance of the analog demapper is evaluated by considering an information-theoretic metric called achievable rate \cite{shannon}. An achievable rate per bit position is the so-called bit-wise mutual information (MI) $I(B_k;L_k)$. Let $B_k\in\{0,1\}$ denote the random bit in position $k$ and $L_k$ its LLR. The MI between $B_k$ and $L_k$ can be approximated via \cite[eq.~(4.82)]{bicm}
\begin{equation}\label{mi}
    I(B_k;L_k) \approx 1 - \mathbb{E}\!\left[\log_2\!\big(1+\exp({(-1)^{B_k} L_k})\big)\right],
\end{equation}
where $\mathbb{E}[\,\cdot\,]$ represents expectation. An achievable rate for the system under consideration with 8-PAM is the average bit-wise MIs, often referred to as the generalized mutual information (GMI) and is expressed as
\begin{equation}\label{gmi}
    \text{GMI} = \frac{1}{3}\sum\limits^{3}_{k=1}I(B_k;L_k).
\end{equation}

In what follows, we will also present results in terms of the rate penalty caused by the approximations. This penalty is defined as the loss caused by max-log or the two analog demappers, in terms of GMI, with respect to the GMI obtained when using exact LLRs. Lastly, bit error rate (BER) is presented by performing hard decisions on the LLRs: $b_k=1$ if $L_k\geq 0$ and $b_k=0$ if $L_k<0$.

A DC sweep of the input voltage was used to obtain the transfer characteristic and corresponding LLRs. Fig.~\ref{fig:llr} shows the LLRs for the exact, max-log, BJT, and MOSFET approximations, indicating that the analog results closely follow the digital references for $k=1$ and $k=2$. For $k=3$ the BJT approximation is still very precise, however, the MOSFET approximation deviates substantially from the exact LLRs. 

\begin{figure}[t]
    \centering
    \definecolor{Cblue}{RGB}{31 119 180}
\definecolor{Cred}{RGB}{214 39 40}
\definecolor{Corange}{RGB}{255 127 14}
\definecolor{Cgreen}{RGB}{44 160 44}
\definecolor{Cpurple}{RGB}{148 103 198}
\definecolor{Cbrown}{RGB}{140 86 75}
\definecolor{Cpink}{RGB}{227 119 194}
\definecolor{Ccyan}{RGB}{23 190 207}
\definecolor{Cgray}{RGB}{127 127 127}
\definecolor{Cgray}{RGB}{0 0 0}
\definecolor{Cyellow}{RGB}{188 189 34}

\begin{tikzpicture}

\begin{axis}[
    width=1.05\columnwidth,
    height=5.5cm,
    xlabel={SNR [$\si{\decibel}$]},
    ylabel={Rate penalty [$\si{\percent}$]},
    ymin=0, ymax=20,
    xmin=-10, xmax=20,
    xticklabel style={/pgf/number format/fixed, font=\footnotesize},
    yticklabel style={font=\footnotesize},
    grid=both,
    thick,
        legend style={
      font=\footnotesize},
    legend pos=north east,
    ylabel style={yshift=-1ex,font=\small},
    xlabel style={yshift=1ex,font=\small},
    legend cell align=left
]

\addplot [color=Cgreen, 
very thick,
mark=square*,
mark options={solid, fill=white},
every mark/.append style={scale=1.1},
mark repeat=2
]
table [x = gamma_SNR_dB, y = AIR_percent] {figures/ML_AIR_percentnew.txt};\addlegendentry{Digital Max-Log};

\addplot [color=Cred, solid,
very thick,
mark=*,
mark options={solid, fill=white},
every mark/.append style={scale=1.2},
mark repeat=2
]
table [x = gamma_SNR_dB, y = AIR_percent] {figures/30u_AIR_percent_every3.txt};\addlegendentry{Analog BJT};

\addplot [color=Cblue, 
very thick,
mark=diamond*,
mark options={solid, fill=white},
every mark/.append style={scale=1.3},
mark repeat=2
]
table [x = gamma_SNR_dB, y = AIR_percent] {figures/10u_AIR_MOS_every3.txt};\addlegendentry{Analog MOSFET};

\end{axis}
\end{tikzpicture}%
    \vspace{-4ex}
    \caption{Rate penalty relative to the digital exact demapper vs SNR of the max-log, BJT and MOSFET approximations}
    \vspace{-2ex}
    \label{fig:air}
\end{figure}
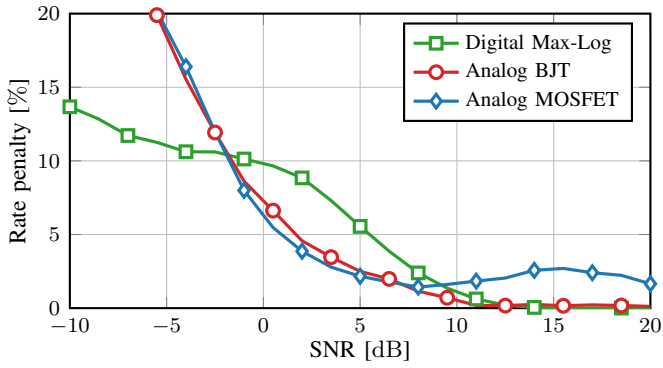

Monte Carlo simulations were performed to evaluate the GMI. The results are shown in Fig.~\ref{fig:air}. The max-log and BJT demappers approach $0\si{\percent}$ penalty at high SNR, while the MOSFET demapper remains within $3\si{\percent}$ for SNRs above $3~\si{\decibel}$ due to the deviation of the LLR for bit position $k=3$ as mentioned earlier. The results in Fig.~\ref{fig:air} also show that when $-1~\si{\decibel}\leq \text{SNR} \leq 9~\si{\decibel}$, the analog demappers outperform the (digital) max-log demapper. These results can be explained by the fact that the first two bit positions ($k=1,2$) are the ones that contribute the most to the GMI in \eqref{gmi}. As shown in Fig.~\ref{fig:llr} (top and middle), the analog demappers approximate the exact LLRs better than the max-log approximation for those bit positions.

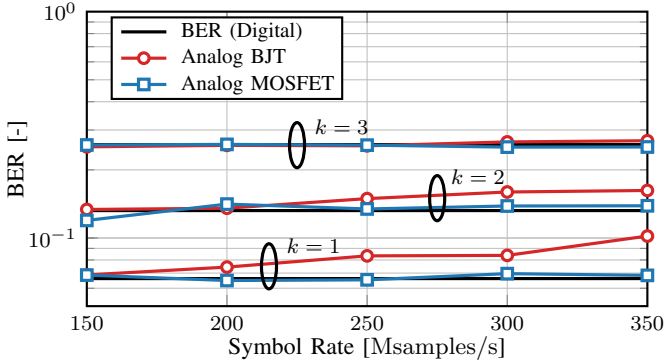
\begin{figure}[t]
    \centering
    \definecolor{Cblue}{RGB}{31 119 180}
\definecolor{Cred}{RGB}{214 39 40}
\definecolor{Corange}{RGB}{255 127 14}
\definecolor{Cgreen}{RGB}{44 160 44}
\definecolor{Cpurple}{RGB}{148 103 198}
\definecolor{Cbrown}{RGB}{140 86 75}
\definecolor{Cpink}{RGB}{227 119 194}
\definecolor{Ccyan}{RGB}{23 190 207}
\definecolor{Cgray}{RGB}{127 127 127}
\definecolor{Cgray}{RGB}{0 0 0}
\definecolor{Cyellow}{RGB}{188 189 34}

\begin{tikzpicture}
\begin{axis}[
    width=1.02\columnwidth,
    height=5.5cm,
    xlabel={Symbol Rate [\si[per-mode=symbol]{Msamples\per\second}]},
    ylabel={BER [-]},
    ymode=log,
    ymin=5e-2, ymax=1,
    xmin=150, xmax=350,
    xtick={150,200,250,300,350},
    xticklabel style={/pgf/number format/fixed, font=\footnotesize},
    yticklabel style={font=\footnotesize},
    grid=both,
    thick,
    clip=false, 
    legend style={
      font=\footnotesize,
      at={(0.25,0.7)},
      anchor=south,
      legend columns=1,
      column sep=5pt,
      nodes={inner sep=0pt},
      row sep=3pt,
      legend cell align=left,
    },
    ylabel style={yshift=-1.5ex,font=\small},
    xlabel style={yshift=1ex,font=\small},
]

\addplot [
    color=black,
    very thick
]
coordinates {
    (150,0.0662)
    (350, 0.0662)
};
\addlegendentry{BER (Digital)}

\addplot [
    color=Cred,
    very thick,
    mark=*,
    mark options={fill=white},
]
coordinates {
    (150,0.0687)
    (200,0.0744)
    (250,0.0834)
    (300,0.0838)
    (350,0.1020)
};
\addlegendentry{Analog BJT}

\addplot [
    color=Cblue,
    very thick,
    mark=square*,
    mark options={solid, fill=white},
]
coordinates {
    (150,0.0686) 
    (200,0.0649) 
    (250,0.06536) 
    (300,0.0696) 
    (350,0.0684) 
};
\addlegendentry{Analog MOSFET}

\addplot [
    color=black,
    very thick
]
coordinates {
    (150,0.1322)
    (350, 0.1322)
};

\addplot [
    color=black,
    very thick
]
coordinates {
    (150,0.2586)
    (350, 0.2586)
};



\addplot [
    color=Cred,
    very thick,
    mark=*,
    mark options={solid, fill=white},
]
coordinates {
    (150,0.1337)
    (200,0.1354)
    (250,0.1493)
    (300,0.1597)
    (350,0.1622)
};

\addplot [
    color=Cred,
    very thick,
    mark=*,
    mark options={solid, fill=white},
]
coordinates {
    (150,0.2528) 
    (200,0.2567) 
    (250,0.2554) 
    (300,0.2658) 
    (350,0.2690) 
};

\addplot [
    color=Cblue,
    very thick,
    mark=square*,
    mark options={solid, fill=white},
]
coordinates {
    (150,0.1196)
    (200,0.1412)
    (250,0.1344)
    (300,0.1385)
    (350,0.1388) 
};
\addplot [
    color=Cblue,
    very thick,
    mark=square*,
    mark options={solid, fill=white},
]
coordinates {
    (150,0.2576)
    (200,0.2592)
    (250,0.2572)
    (300,0.2521)
    (350,0.2524)
};

\draw [black, very thick] (axis cs:225,0.2572) ellipse (0.1cm and 0.3cm) node[right, font=\footnotesize, text=black, xshift=0.1cm, yshift=7pt] {$k=3$};

\draw [black, very thick] (axis cs:275,0.15) ellipse (0.1cm and 0.3cm) node[right, font=\footnotesize, text=black, xshift=0.05cm, yshift=8pt] {$k=2$};

\draw [black, very thick] (axis cs:215,0.075) ellipse (0.1cm and 0.3cm) node[right, font=\footnotesize, text=black, xshift=0.1cm, yshift=8pt] {$k=1$};

\end{axis}
\end{tikzpicture}
    \vspace{-4ex}
    \caption{BER performance over sampling frequency. The horizontal lines indicate the reference BER from hard decisions on LLRs from the digital exact demapper at $\text{SNR}=10~\si{\decibel}$, representing the best achievable performance.}
    \label{fig:transient}
\end{figure}

\begin{figure}
    \centering
    
    \begin{subfigure}{0.49\linewidth}
        \centering
        \definecolor{Cblue}{RGB}{31 119 180}
\definecolor{Cred}{RGB}{214 39 40}
\definecolor{Corange}{RGB}{255 127 14}
\definecolor{Cgreen}{RGB}{44 160 44}
\definecolor{Cpurple}{RGB}{148 103 198}
\definecolor{Cbrown}{RGB}{140 86 75}
\definecolor{Cpink}{RGB}{227 119 194}
\definecolor{Ccyan}{RGB}{23 190 207}
\definecolor{Cgray}{RGB}{127 127 127}
\definecolor{Cgray}{RGB}{0 0 0}
\definecolor{Cyellow}{RGB}{188 189 34}

\begin{tikzpicture}

\begin{axis}[
    width=5.2cm,
    height=3.8cm,
    xlabel={$t$~$[\si{\nano\second}]$},
    ylabel = {$V_\text{out}~[\si{\volt}]$},
    xmin=0, xmax=10,
    xticklabel style={/pgf/number format/fixed, 
    font=\footnotesize},
    ytick={-0.05,-0.06,-0.07,-0.08,-0.09},
    yticklabel style={font=\footnotesize},
    grid=both,
    thick,
        legend style={
      font=\scriptsize},
    legend pos=north east,
    legend cell align=left,
    xlabel style={yshift=1ex,font=\small},
    ylabel style={yshift=-1.5ex,font=\small},
]

\addplot [color=Cred, 
very thick]
table [x expr=\thisrow{time}*1e9, y expr = \thisrow{vOut}, col sep=space] {figures/Transitiondatabjt/group_56_segment_170.txt};\addlegendentry{\footnotesize BJT};

\addplot [color=Cblue, 
very thick]
table [x expr=\thisrow{time}*1e9, y expr = (\thisrow{vOut} - 0.0164266), col sep=space] {figures/Transitiondatamos/group_56_segment_18.txt};\addlegendentry{\footnotesize MOSFET};

\end{axis}
\end{tikzpicture}%
    \end{subfigure}
    \hfill
    \begin{subfigure}{0.49\linewidth}
        \centering
        \definecolor{Cblue}{RGB}{31 119 180}
\definecolor{Cred}{RGB}{214 39 40}
\definecolor{Corange}{RGB}{255 127 14}
\definecolor{Cgreen}{RGB}{44 160 44}
\definecolor{Cpurple}{RGB}{148 103 198}
\definecolor{Cbrown}{RGB}{140 86 75}
\definecolor{Cpink}{RGB}{227 119 194}
\definecolor{Ccyan}{RGB}{23 190 207}
\definecolor{Cgray}{RGB}{127 127 127}
\definecolor{Cgray}{RGB}{0 0 0}
\definecolor{Cyellow}{RGB}{188 189 34}

\begin{tikzpicture}

\begin{axis}[
    width=5.2cm,
    height=3.8cm,
    xlabel={$t$~[\si{\nano\second}]},
    xmin=0, xmax=5,
    xticklabel style={/pgf/number format/fixed, font=\footnotesize},
    ytick = {0.09,0.08,0.07,0.06,0.05},
    xtick={0,1,2,3,4,5},
    yticklabel style={font=\footnotesize},
    ymax = 0.1,
    grid=both,
    thick,
    legend pos=north east,
    legend cell align=left,
    xlabel style={yshift=1ex,font=\small},
    scaled y ticks = base 10:2,
]

\addplot [color=Cred, 
very thick]
table [x expr=\thisrow{time}*1e9, y = vOut, col sep=space] {figures/Transitiondatabjt/group_02_segment_59.txt};\addlegendentry{\footnotesize BJT};

\addplot [color=Cblue, 
very thick]
table [x expr=\thisrow{time}*1e9, y expr = \thisrow{vOut} + 0.0131162, col sep=space] {figures/Transitiondatamos/group_02_segment_226.txt};\addlegendentry{\footnotesize MOSFET};

\end{axis}
\end{tikzpicture}%
    \end{subfigure}
    \vspace{-4ex}
    \caption{Comparison of the settling behavior for two symbol transitions in the BJT and MOSFET demappers for $k=1$: Transition from $r=+3d$ to $r=+7d$ (left) and from $r=-7d$ to $r=-5d$ (right). A vertical offset has been applied to the MOSFET waveforms to allow comparison.
    }
    \vspace{-2ex}
    \label{fig:eyediag}
\end{figure}
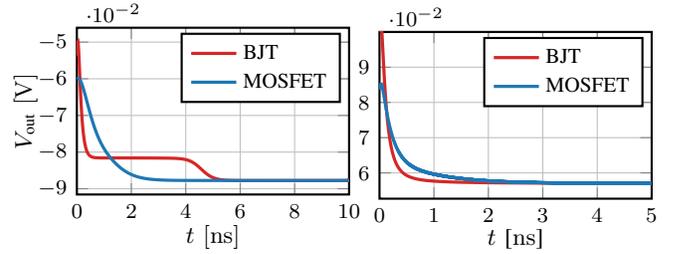

A transient simulation was performed to evaluate demapping speed. The input of the analog demapper are voltage levels of a sequence of square pulses of varying symbol rate. The output of the analog demapper was sampled, which allowed us to assess the circuit’s dynamic response and settling behavior. The resulting BER performance is shown in Fig.~\ref{fig:transient}. From Fig.~\ref{fig:transient}, it can be seen that the BJT demapper exhibits an increasing BER as the symbol rate rises, while the MOSFET design maintains a constant BER across the entire symbol rate range, very close to that of the ideal (digital) demapper. This behavior can be further understood from Fig.~\ref{fig:eyediag} (left), which illustrates the output voltage dynamics for a symbol transition that requires certain cells to be driven out of saturation. As the BJTs exit the saturation region, they must first discharge their base, resulting in the plateau visible in the red waveform. The MOSFET, however, does not show such a plateau, and thus, settles much faster (after $t\approx 2~\si{\nano\second}$). On the other hand, Fig.~\ref{fig:eyediag} (right) corresponds to a transition where all relevant BJT cells are in the forward-active region, therefore no BJT needs to come out of saturation. In this case, no plateau appears in the BJT response, showing that if BJTs remain out of saturation, the demapper based on BJTs can achieve settling behavior comparable to that of the MOSFET circuit.

To obtain the results shown in Fig.~\ref{fig:transient}, the BJT and MOSFET designs were biased at $I_\text{bias}=100~\si{\micro\ampere}$ and $I_\text{bias}=10~\si{\micro\ampere}$ per cell, resp. Reducing the bias current in the BJT design led to a significant increase in BER, and thus, a relatively high bias current had to be used ($I_\text{bias}=100~\si{\micro\ampere}$). Our MOSFET design on the other hand, offers a near-optimum BER performance for symbol rates up to $350~\si{Msamples/\second}$, which corresponds to a bit rate of about $1~\si{\giga\bit/\second}$. The power consumption of our design is $0.35~\si{\milli\watt}$, i.e., $0.33~\si{\pico\joule/\bit}$ at $1~\si{\giga\bit/\second}$.

\section{Conclusions}\label{sec.conc}
This paper presented a fully open-source implementation of an analog 8-PAM demapper. The proposed design achieves an energy efficiency of $0.33~\si{\pico\joule/\bit}$ for a data rate of $1~\si{\giga\bit/s}$. Compared to digital demappers, the proposed solution provides analog LLRs that can be directly interfaced with an analog FEC decoder. The results highlight a trade-off between accuracy and speed across BJT and MOSFET implementations and confirm the potential of efficient analog demappers for high-speed communication systems.

The results in this paper are simple proof-of-concept results highlighting the capabilities of the open-source design flow based on IHP-Open-PDK and open-source EDA tools. Future research includes an analysis of analog demappers for denser, geometrically-shaped constellations. Future work also includes PVT simulations and an analysis of the impact of circuit noise on the design.

\newpage
\bibliographystyle{IEEEtran}
\bibliography{sample.bib}
\end{document}